\documentstyle[epsfig]{aipproc}
\newcommand{\beq}{\begin{equation}} 
\newcommand{\eeq}{\end{equation}} 
\newcommand{\barr}{\begin{eqnarray}} 
\newcommand{\earr}{\end{eqnarray}} 
\newcommand{\ba}{\begin{array}} 
\newcommand{\ea}{\end{array}} 
\newcommand{\bfp}{\mbox{\boldmath $p$}} 
\newcommand{\bfP}{\mbox{\boldmath $P$}} 
\newcommand{\bfk}{\mbox{\boldmath $k$}}

\newcommand{\pup}{p^\uparrow}

\newcommand{\qup}{q^\uparrow}

\newcommand{\Lup}{\Lambda^\uparrow} 
\newcommand{\Ldown}{\Lambda^\downarrow}

\newcommand{\hup}{h^\uparrow} 
 
\newcommand{\simorder}{\raisebox{-4pt}{$\, \stackrel{\textstyle >}{\sim} \,$}} 
\newcommand{\NP}[1]{{\it Nucl.\ Phys.}\ {\bf #1}} 
 
\newcommand{\EPJ}[1]{{\it Eur. Phys. J.}\ {\bf #1}} 
\newcommand{\PL}[1]{{\it Phys.\ Lett.}\ {\bf #1}} 
\newcommand{\PR}[1]{{\it Phys.\ Rev.}\ {\bf #1}} 
\newcommand{\PRL}[1]{{\it Phys.\ Rev.\ Lett.}\ {\bf #1}} 
\newcommand{\IJMP}[1]{{\it Int.\ J.\ Mod.\ Phys.}\ {\bf #1}}

\begin{document}
%%%%%%%%%%%%%%%%%%%%%%%%%%%%%%%%%%%%%%%%%%%%%%%%%%%%%%%%%%%%%%%%%%%%%%%%%%%%%% 
%\begin{flushright} 
%DFTT 32/2000 \\ 
%INFNCA-TH0012 \\ 
%hep-ph/0011361 \\ 
%\end{flushright} 
%%%%%%%%%%%%%%%%%%%%%%%%%%%%%%%%%%%%%%%%%%%%%%%%%%%%%%%%%%%%%%%%%%%%%%%%%%%%%% 
\renewcommand{\thefootnote}{\fnsymbol{footnote}}
\title{Spin effects in the fragmentation of transversely polarized 
and unpolarized quarks\footnote{Talk delivered by M. Anselmino
at the ``14th International Spin Physics Symposium'', SPIN2000,
October 16-21, 2000, Osaka, Japan.}}
\author{M. Anselmino$^a$, D. Boer$^b$, U. D'Alesio$^c$
 and F. Murgia$^c$}
\address{$^a$Dipartimento di Fisica Teorica, Universit\`a di Torino and \\ 
      INFN, Sezione di Torino, Via P. Giuria 1, I-10125 Torino, Italy\\ 
\vskip 0.2cm 
$^b$ RIKEN-BNL Research Center\\ 
Brookhaven National Laboratory, Upton, NY 11973, USA\\  
\vskip 0.2cm 
$^c$  Istituto Nazionale di Fisica Nucleare, Sezione di Cagliari\\ 
      and Dipartimento di Fisica, Universit\`a di Cagliari\\ 
      C.P. 170, I-09042 Monserrato (CA), Italy} 
%\lefthead{LEFT head}
%\righthead{RIGHT head}
\maketitle
\begin{abstract}
We study the fragmentation of a transversely polarized quark
into a non collinear ($\bfk_\perp \not= 0$) spinless hadron and the 
fragmentation of an unpolarized quark into a non collinear transversely 
polarized spin 1/2 baryon. These nonperturbative properties are 
described by spin and $\bfk_\perp$ dependent fragmentation functions
and are revealed in the observation of single spin asymmetries.
Recent data on the production of pions in polarized semi-inclusive DIS 
and long known data on $\Lambda$ polarization in unpolarized 
$p$--$N$ processes are considered: these new fragmentation functions can 
describe the experimental results and the single spin effects in the quark 
fragmentation turn out to be surprisingly large.
\end{abstract}

\section*{Introduction}
Several large and puzzling single spin asymmetries in high energy inclusive 
processes are experimentally well known since a long time and new ones
have just been or are being measured. These effects are absent in massless
perturbative QCD dynamics and they depend on new and interesting aspects 
of nonperturbative QCD; as such, they deserve a careful study, both 
theoretically and experimentally.

We consider here spin properties of quark fragmentation processes,
which have been suggested in the literature \cite{col,multan,noi1}, 
and address the question of
whether or not they might explain some single spin asymmetries observed in
inclusive processes: in particular we look at the recently observed 
azimuthal dependence of the number of pions produced in polarized 
semi-inclusive DIS, $\ell \pup \to \ell \pi X$ \cite{her,smc}, 
and at the longstanding problem of the polarization of $\Lambda$'s produced 
in unpolarized $p$--$p$ and $p$--$n$ interactions, $p N \to \Lup X$ 
\cite{lam}. Both these unexpected spin dependences should originate in the 
fragmentation process of a quark, polarized in the first case ($\qup \to
\pi X$) and unpolarized in the second one ($q \to \Lup X$).
     
\section*{Quark analysing power}
The inclusive production of hadrons in DIS with transversely polarized
nucleons, $\ell N^\uparrow \to \ell h X$, is the ideal process to study 
the so-called Collins effect, {\it i.e.} the spin and $\bfk_\perp$ 
dependence of the fragmentation process of a transversely polarized quark,
$\qup \to hX$. In such a case, in fact, possible effects in quark distribution
functions \cite{siv}, which require initial state interactions 
\cite{noi2,noi3}, are expected to be negligible.

If one looks at the $\gamma^* N^\uparrow \to h X$ process in the 
$\gamma^*$--$N$ c.m. frame, the elementary interaction is simply a 
$\gamma^*$ hitting head on a transversely polarized quark, which bounces 
back and fragments into a jet containing the detected hadron. The hadron 
$\bfp_T$ in this case coincides with its $\bfk_\perp$ inside the jet; the 
fragmenting quark polarization can be computed from the initial quark one. 

The spin and $\bfk_\perp$ dependent fragmentation function for a quark with 
momentum $\bfp_q$ and a {\it transverse} polarization vector $\bfP_q$ 
($\bfp_q \cdot \bfP_q = 0$) which fragments into a hadron with momentum 
$\bfp_h = z\bfp_q + \bfp_T$ ($\bfp_q \cdot \bfp_T = 0$) can be written as:
\beq 
D_{h/q}(\bfp_q, \bfP_q; z, \bfp_T) = \hat D_{h/q}(z, p_T) + \frac 12 \> 
\Delta^ND_{h/\qup}(z, p_T) \> \frac{\bfP_q \cdot (\bfp_q \times \bfp_T)}
{|\bfp_q \times \bfp_T|} \label{colfn}
\eeq
where $\hat D_{h/q}(z, p_T)$ is the unpolarized fragmentation function. 
Notice that -- as required by parity invariance -- the only component 
of the polarization vector which contributes to the spin dependent part
of $D$ is that perpendicular to the $q-h$ plane; in general one has:
\beq
\bfP_q \cdot 
\frac{\bfp_q \times \bfp_T} {|\bfp_q \times \bfp_T|}
= P_q \sin\Phi_C \>, \label{colan}
\eeq
where $P_q = |\bfP_q|$ and we have defined the {\it Collins angle} $\Phi_C$. 

When studying single spin asymmetries one considers differences of 
cross-sections with opposite transverse spins; by reversing the nucleon 
spin all polarization vectors, including those of quarks, change sign 
and the quantity which eventually contributes to single spin asymmetries is:
\beq 
D_{h/q}(\bfp_q, \bfP_q; z, \bfp_T) - D_{h/q}(\bfp_q, -\bfP_q; z, \bfp_T)
= \Delta^ND_{h/\qup}(z, p_T) \> \frac{\bfP_q \cdot (\bfp_q \times \bfp_T)}
{|\bfp_q \times \bfp_T|} \label{coldf}
\eeq
which implies the existence of a {\it quark analysing power}
for the fragmentation process $q \to h + X$:
\barr
A_q^h (\bfp_q, \bfP_q; z, \bfp_T) &=& \frac 
{D_{h/q}(\bfp_q, \bfP_q; z, \bfp_T) - D_{h/q}(\bfp_q, -\bfP_q; z, \bfp_T)}
{D_{h/q}(\bfp_q, \bfP_q; z, \bfp_T) + D_{h/q}(\bfp_q, -\bfP_q; z, \bfp_T)}
\label{aq} \\
&=&\frac {\Delta^ND_{h/\qup}(z, p_T)}   
{2\,\hat D_{h/q}(z, p_T)} \> \frac {\bfP_q \cdot (\bfp_q \times \bfp_T)}
{|\bfp_q \times \bfp_T|} 
\equiv A_q^h (z, p_T) \> \frac {\bfP_q \cdot (\bfp_q \times \bfp_T)}
{|\bfp_q \times \bfp_T|} \> \cdot \nonumber 
\earr

This results in a single spin asymmetry \cite{noi4}:
\barr  
A^h_N(x,y,z,\Phi_C,p_T) &=&
 \frac{d\sigma^{\ell + p,\bfP \to \ell + h + X}
      -d\sigma^{\ell + p,-\bfP \to \ell + h + X}}
      {d\sigma^{\ell + p,\bfP \to \ell + h + X}
      +d\sigma^{\ell + p,-\bfP \to \ell + h + X}} \nonumber \\
&=&\frac{\sum_q e_q^2 \, h_{1q}(x) \> \Delta^ND_{h/q}(z, p_T)}
{2\sum_q e_q^2 \, f_{q/p}(x) \> \hat D_{h/q}(z, p_T)} \>
\frac{2(1-y)} {1 + (1-y)^2} \> P \> \sin\Phi_C \>, \label{asym1} 
\earr
where $P$ is the transverse (with respect to the $\gamma^*$ direction)
proton polarization.

We wonder how large the quark analysing power can be. Such a question
has been addressed in Ref. \cite{noi4}, where recent data on 
$A^\pi_N$ \cite{her,smc} were considered. We refer to that paper
for all the details and only outline the main procedure here. 
Under some realistic assumptions and using isospin and charge 
conjugation invariance Eq. (\ref{asym1}) gives ($i = +,-,0$):
\beq
A^{\pi^i}_N(x,y,z,\Phi_C, p_T) = 
\frac{h_i(x)}{f_i(x)} \> A_q^\pi(z, p_T) \> 
\frac{2(1-y)} {1 + (1-y)^2} \> P \> \sin\Phi_C \label{aspi} 
\eeq
where:
\barr
i &=& + : \quad h_+ = 4h_{1u} \quad\quad 
f_+ = 4f_{u/p} +  f_{\bar d/p} \label{hf+} \\ 
i &=& - : \quad h_- =  h_{1d} \quad\quad 
f_- =  f_{d/p} + 4f_{\bar u/p}  \label{hf-} \\ 
i = 0 : \quad h_0 &=& 4h_{1u} + h_{1d} \quad\quad 
f_0 = 4f_{u/p} + f_{d/p} + 4f_{\bar u/p} + f_{\bar d/p} \>. \label{hf0}
\earr
  
The $f$'s are the unpolarized distribution functions and the $h_1$'s
are the transversity distributions. Notice that the above equations
imply -- at large $x$ values -- $A^{\pi^+}_N \simeq A^{\pi^0}_N$ as 
observed by HERMES \cite{talk}.

We bound the unknown transversity distributions saturating Soffer's 
inequality \cite{sof}
\beq
|h_{1q}| \le \frac12 \, (f_{q/p} + \Delta q) \>, \label{sofb}
\eeq
and, by comparing with SMC data \cite{smc}
\beq
A_N^{\pi^+} \simeq -(0.10 \pm 0.06)\,\sin\Phi_C \>, \label{an+}
\eeq
we obtain the significant lower bound for pion valence quarks:
\beq
|A_q^\pi(\langle z \rangle, \langle p_T \rangle)| \>
\simorder \> (0.24 \pm 0.15)
\quad\quad \langle z \rangle \simeq 0.45 \>, \quad 
\langle p_T \rangle \simeq 0.65 \> \mbox{GeV}/c \>.
\label{res}
\eeq  

A similar result is obtained by using HERMES data \cite{her}, although their
transverse polarization is much smaller. These lower bounds of the
quark analysing power are remarkably large and indeed the Collins
mechanism might be (at least partly) responsible for several other observed
single transverse spin asymmetries \cite{noi2}.

\section*{Quark polarizing fragmentation functions}
We consider now the possibility that an unpolarized quark fragments into 
a transversely polarized hadron \cite{multan,noi1}; in analogy to 
Eq. (\ref{colfn}) we write
\beq  
\hat D_{\hup/q}(z, \bfk_\perp) = \frac 12 \> \hat D_{h/q}(z, k_\perp) +  
\frac 12 \> \Delta^ND_{\hup/q}(z, k_\perp) \>  
\frac{\hat{\bfP}_h \cdot (\bfp_q \times \bfk_\perp)} 
{|\bfp_q \times \bfk_\perp|} \label{lamfn} 
\eeq 
for an unpolarized quark with momentum $\bfp_q$ which fragments into  
a spin 1/2 hadron $h$ with momentum $\bfp_h = z \bfp_q + \bfk_\perp$ 
and polarization vector along the $\uparrow \> = \hat{\bfP}_h$ direction. 
$\Delta^ND_{\hup/q}(z, k_\perp)$ (denoted by $D_{1T}^\perp$ in 
Ref. \cite{multan}) is a new {\it polarizing fragmentation function}.

This reflects into a possible polarization of hadrons inclusively
produced in the high energy interaction of unpolarized nucleons. 
Indeed, it is well known since a long time that $\Lambda$ hyperons 
produced with $x_F \simorder 0.2$ and $p_T \simorder$ 1 GeV/$c$ in the 
collision of two unpolarized nucleons are polarized perpendicularly to 
the production plane, as allowed by parity invariance; a huge amount of 
experimental information, for a wide energy range of the unpolarized beams, 
is available on such single spin asymmetries \cite{lam}: 
\beq 
P_\Lambda =  
\frac{d\sigma^{p N \to \Lup X} - d\sigma^{p N \to \Ldown X}} 
     {d\sigma^{p N \to \Lup X} + d\sigma^{p N \to \Ldown X}} \>\cdot 
\label{pol} 
\eeq 

By taking into account intrinsic $\bfk_\perp$ in the hadronization  
process, and assuming that the factorization theorem holds also when  
$\bfk_\perp$'s are included \cite{col}, one obtains
\barr 
\frac{E_\Lambda \, d^3\sigma^{pN \to \Lambda X}}{d^3 \bfp_\Lambda} \>  
P_\Lambda &=& 
\sum_{a,b,c,d} \int \frac{dx_a \, dx_b \, dz}{\pi z^2} \> d^2\bfk_\perp \>  
f_{a/p}(x_a) \> f_{b/N}(x_b) \nonumber \\ 
&\times& \hat s \> \delta(\hat s + \hat t + \hat u) \>  
\frac{d\hat\sigma^{ab \to cd}}{d\hat t}(x_a, x_b, \bfk_\perp) \>  
\Delta^ND_{\Lup/c}(z, \bfk_\perp) \label{phgen} 
\earr 

We use the above equation, together with a simple parameterization
of the polarizing fragmentation functions, to see whether or not one can
fit the experimental data on $\Lambda$ and $\bar \Lambda$ polarization. 
All details can be found in Ref. \cite{noi1} and some results are shown 
here in Fig. 1.

\begin{figure}[t]
\begin{center}
  \epsfig{file=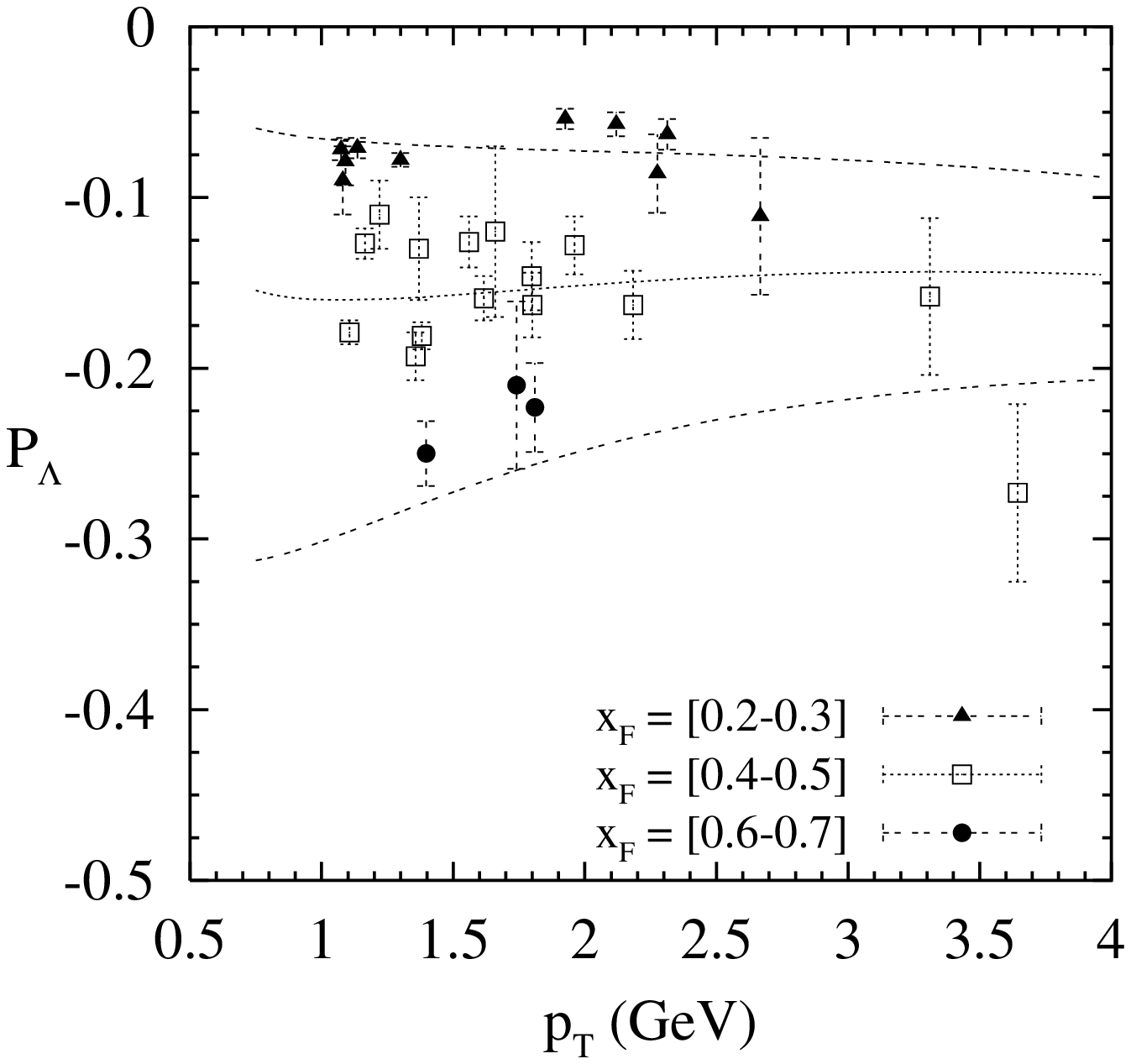,angle=-90,width=7.3cm} 
  \epsfig{file=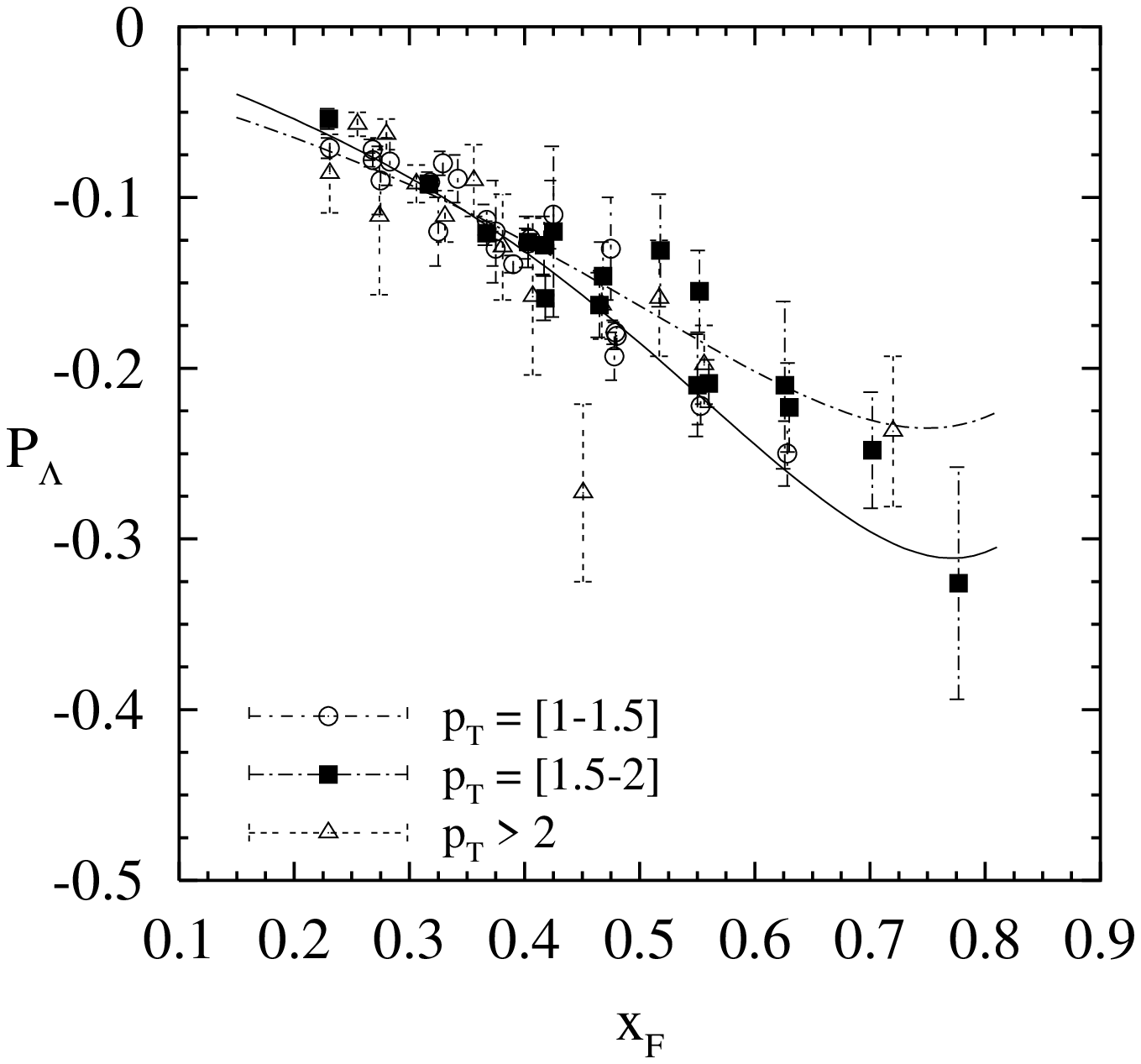,angle=-90,width=7.3cm} 
\end{center}
\vspace{-2mm}
\caption{
   Our best fit to $P_\Lambda$ data  from $p$--$Be$ reactions 
   (a partial collection from  Ref.~\protect\cite{lam}) 
   as a function of $p_T$ (on the left) and of $x_F$ (on the right). 
   For each $x_F$-bin, the corresponding theoretical curve is evaluated 
   at the mean $x_F$ value in the bin.
   The two theoretical curves, on the right, 
   correspond to $p_T=1.5$ GeV$/c$ (solid) and $p_T=3$ GeV$/c$ 
   (dot-dashed). 
} 
\end{figure}

The data can be described with remarkable accuracy in all their features:
the large negative values of the $\Lambda$ polarization, the increase of
its magnitude with $x_F$, the puzzling flat $p_T \simorder 1$ GeV/$c$ 
dependence and the $\sqrt s$ apparent independence; data from $p$--$p$
processes are in agreement with data from $p$--$Be$ interactions and also the 
tiny or zero values of $\bar\Lambda$ polarization are well reproduced. 
The resulting functions $\Delta^ND_{\Lup/q}$ are very reasonable and 
realistic.  

We conclude by stressing that a systematic phenomenological approach 
towards the description and prediction of single transverse spin asymmetries, 
based on perturbative QCD dynamics and nonperturbative quark properties, is
now possible and worth being developed.

\end{document}